\documentclass[a4paper,prd,twocolumn,preprintnumbers,superscriptaddress,nofootinbib,amsmath,amssymb]{revtex4-2}
\usepackage[dvips]{graphics}
\usepackage{color,hyperref}
\usepackage{times}
\usepackage{mathrsfs}
\hypersetup{colorlinks=true,linkcolor=blue,citecolor=red,filecolor=magenta,urlcolor=blue}
\usepackage{enumitem}
\usepackage{graphicx,subfigure}
\usepackage{dcolumn}
\usepackage{bm}
\usepackage{color}
\usepackage{babel}

\begin{document}

\title{Electromagnetic quasinormal modes of a Schwarzschild black hole in a superconducting plasma}

\author{Bobur Turimov}
\email{bturimov@astrin.uz}
\affiliation{School of Physics, Harbin Institute of Technology, Harbin 150001, People's Republic of China}
\affiliation{Central Asian University, Milliy bog Str. 264, Tashkent 111221, Uzbekistan}
\affiliation{University of Tashkent for Applied Sciences, Gavhar Str. 1, Tashkent 100149, Uzbekistan}

\author{Akbar Davlataliev}
\email{akbar@astrin.uz}
\affiliation{School of Physics, Harbin Institute of Technology, Harbin 150001, People's Republic of China}
\affiliation{New Uzbekistan University, Movarounnahr str. 1, Tashkent 100000, Uzbekistan}

\author{Bobomurat Ahmedov}
\email{ahmedov@astrin.uz}
\affiliation{School of Physics, Harbin Institute of Technology, Harbin 150001, People's Republic of China}
\affiliation{Institute of Theoretical Physics, National University of Uzbekistan, Tashkent 100174, Uzbekistan}

\author{Chengxun Yuan}
\email{yuancx@hit.edu.cn}
\affiliation{School of Physics, Harbin Institute of Technology, Harbin 150001, People's Republic of China}

\author{Chen Zhou}
\email{chenzhou@hit.edu.cn}
\affiliation{School of Physics, Harbin Institute of Technology, Harbin 150001, People's Republic of China}

\author{Sulton Usanov}
\email{sm.usanov@kiut.uz}
\affiliation{Kimyo International University in Tashkent, Shota Rustaveli street 156, Tashkent 100121, Uzbekistan}

\date{\today}

\begin{abstract}
We formulate a generally covariant extension of London electrodynamics in four-dimensional curved spacetime and investigate the propagation of electromagnetic perturbations in a superconducting medium surrounding a black hole. The ordinary London relation is promoted to a gauge-invariant four-dimensional constitutive equation involving the electromagnetic four-potential and the phase of the superconducting condensate. When combined with the covariant Maxwell equations, the resulting system is equivalent, in the London (unitary) gauge, to a Proca theory with an effective photon mass set by the inverse London penetration depth. We derive the corresponding Schr\"odinger-like wave equations for electromagnetic perturbations in the axial and polar sectors of the Schwarzschild spacetime in the presence of the superconducting plasma. The axial sector is governed by a single Regge--Wheeler-type equation whose effective potential contains the London mass term, whereas the polar sector contains two coupled physical degrees of freedom. Quasinormal-mode boundary conditions are formulated for the Schwarzschild geometry. The London penetration depth shifts the oscillation frequencies, modifies the damping rates, and can generate long-lived quasibound configurations when the effective field mass is sufficiently large. This formalism provides a framework for studying electromagnetic relaxation in curved superconducting media, compact objects containing superconducting phases, and effective massive-vector fields around black holes.
\end{abstract}

\maketitle

\section{Introduction}

Black holes are among the most fascinating and extreme objects predicted by general relativity and serve as unique laboratories for probing the intersection of gravity, quantum mechanics, and thermodynamics. A central tool in understanding their physical properties is the study of their response to perturbations. The characteristic quasinormal modes (QNMs)---damped oscillations that dominate the ringdown phase of a perturbed black hole---provide a direct link between theoretical predictions~\cite{Berti:2009wt,Kokkotas:1999bd,Konoplya:2011qq,Nollert:1999CQG} and the gravitational-wave signals observed by LIGO and Virgo~\cite{LIGOVirgo2016}. However, the vacuum description of a black hole is an idealization. In realistic astrophysical scenarios, black holes are rarely isolated; they are often embedded in complex environments such as accretion disks, molecular clouds, or, as considered here, a plasma~\cite{Rees:1984zz,Frank:2002zz}. The analytical approach to black-hole quasinormal modes has been developed in~\cite{Schutz:1985ApJL,Iyer:1987PRD,Leaver1985}. The structure and dynamics of an accretion disk are strongly influenced by the spacetime geometry surrounding the black hole. In particular, quantities such as the radius of the innermost stable circular orbit, the orbital energy and angular momentum of disk particles, the radiative efficiency, and the disk temperature depend on the mass, angular momentum, electric charge, and possible additional parameters of the black-hole spacetime. Therefore, observations of radiation emitted from accretion disks can be used to test the predictions of general relativity and to constrain alternative theories of gravity~\cite{Pani2012}.

The presence of a medium around a black hole fundamentally alters the propagation of fields. In particular, a plasma can dramatically affect the behavior of electromagnetic (EM) waves, leading to phenomena such as dispersion, absorption, and reflection~\cite{Breuer:1975zz,Holcomb:1998zz}. The propagation of electromagnetic waves in a plasma environment around a strong Schwarzschild black hole was studied in~\cite{Daniel1997}. A particularly interesting and physically rich case is that of a superconducting plasma~\cite{Tinkham1996,Salasnich2024}. The landmark discovery of the Meissner effect~\cite{Meissner1933,Bicak2000,Penna2014} and the subsequent formulation of London electrodynamics revealed that photons acquire an effective mass inside a superconductor, leading to the screening of magnetic fields~\cite{London1935,Tinkham:1996zz}. This effective mass substantially changes the propagation of electromagnetic disturbances in the vicinity of a black hole. In ordinary vacuum electrodynamics, Maxwell perturbations are massless and possess two transverse physical polarizations. By contrast, a massive vector field has three propagating degrees of freedom, including an additional longitudinal polarization. The presence of this longitudinal mode modifies the perturbation structure, the asymptotic behavior of the field, and the spectrum of characteristic oscillations. Consequently, superconducting electromagnetic perturbations near a black hole cannot generally be described by the standard Maxwell quasinormal-mode equations. The effective mass generation is a fundamental example of a spontaneous symmetry-breaking mechanism, conceptually similar to the Higgs mechanism in particle physics~\cite{Anderson:1963zz,Higgs:1964zz}.

Bounds on electric charge derived from gravitational-wave dephasing during the inspiral of two compact objects, or from black-hole spectroscopy, rest on an implicit assumption: that photons propagate freely from the source to the observer~\cite{Khalil:2018PRD,Carullo:2021PRD}. The Universe, however, is not empty. Even a dilute interstellar plasma prevents electromagnetic waves below the plasma frequency from propagating, behaving effectively as a photon mass~\cite{Cardoso:2020mnras}
\begin{align}
\omega_p=\sqrt{\frac{e^2n_e}{m_e}}\simeq 1.8\times 10^3\left(\frac{n_e}{10^{-3}{\rm cm}}\right)^{1/2}{\rm rad/s}    
\end{align}
where $n_e$ is the plasma number density. On the other hand, the QNM's frequency of the stellar Schwarzschild black hole is
\begin{align}
\omega_{\rm QNM}=\frac{c^3}{GM}\simeq 2\times 10^4\left(\frac{M}{10M_\odot}\right)^{-1}{\rm rad/s}\ .    
\end{align}

The standard London equations are not manifestly covariant, and their naive extension to curved spacetime can lead to inconsistencies with fundamental principles such as gauge invariance~\cite{Dewitt:1965zz,Ahmedov:1999bqr}. By promoting the constitutive relation to a four-dimensional equation involving the electromagnetic four-potential and the phase of the superconducting condensate, we ensure that the theory is consistent with the principles of general relativity and electromagnetism. This covariant formulation reveals a profound and elegant connection: in the appropriate gauge, the theory of a superconductor is equivalent to a Proca theory describing a massive vector field~\cite{Proca:1936zz,Ryder:1996zz}. The effective mass of the photon is directly determined by the inverse London penetration depth, which in turn depends on the properties of the plasma. This insight allows us to systematically investigate the perturbations of the coupled Einstein--Maxwell--Proca system. The decomposition of these perturbations into axial (odd-parity) and polar (even-parity) sectors is crucial for understanding the distinct physical degrees of freedom involved~\cite{Regge:1957zz,Zerilli:1970zz,Chandrasekhar:1985kt}. The conformal invariance of Maxwell's equations has been studied in~\cite{Cote:2019GRG}.

Black-hole perturbation theory provides a natural framework for studying the dynamical response of the London--Maxwell field. One begins by treating the superconducting electromagnetic field as a linear perturbation propagating on a fixed black-hole background. For a static and spherically symmetric spacetime, the vector perturbation may be decomposed into spherical harmonics and separated into odd- and even-parity sectors. This decomposition is particularly useful because perturbations belonging to different parity sectors evolve independently at the linear level. The even-parity, or polar, sector is more involved because the temporal, radial, and angular components of the massive vector field are coupled through the field equations and the Lorenz-type constraint. After eliminating the nondynamical variables, the polar sector can be written as a system of two coupled radial wave equations. These equations describe the interaction between the transverse electric-type polarization and the additional longitudinal degree of freedom. In the massless limit, the longitudinal mode disappears, and the system reduces to the familiar electromagnetic perturbation equations. For a nonzero London mass, however, the axial and polar spectra are generally no longer isospectral. The connection between black-hole bombs and photon-mass bounds has been studied in~\cite{Pani2012}. Energy extraction, black-hole bombs, and their implications for astrophysics and particle physics have been studied in~\cite{Brito2015}.

The quasinormal-mode spectrum of the covariant London--Maxwell system can therefore serve as a probe of the interaction between superconducting media and strong gravitational fields. In particular, the frequencies and damping times encode information about the London penetration depth, the effective photon mass, the background spacetime geometry, and the polarization structure of the perturbation. Deviations from the standard Maxwell spectrum may provide a theoretical signature of superconducting or charged-condensate effects in the environment surrounding a compact object~\cite{Cannizzaro2024}. The Schwarzschild radial functions are extensively treated in the appendices, and numerical results are presented for various frequencies and radii. A simple asymptotic expansion for one of the connection constants, appropriate for high frequency, is also given~\cite{Kearney1978}. For rotating black holes, the perturbation problem is technically more complicated because spherical symmetry is lost and the axial--polar decomposition is no longer directly applicable. Nevertheless, massive-vector field equations are separable in important classes of rotating black-hole spacetimes. The separated angular and radial equations determine the quasinormal and quasibound spectra as functions of the black-hole mass, angular momentum, multipole numbers, polarization state, and effective London mass. The rotation of the black hole also splits modes that are degenerate in the nonrotating limit and introduces a dependence on the azimuthal number~\cite{Frolov:2018PRL,Percival:2020PRD}.

Our analysis, as detailed in this paper, shows that the axial sector is governed by a single modified Regge--Wheeler equation whose effective potential includes a new term arising from the photon mass. In contrast, the polar sector is more complex, containing two coupled degrees of freedom that reflect the longitudinal polarization mode of the massive photon~\cite{Rosa:2012zz}. By deriving the corresponding Schr\"odinger-like wave equations for both sectors, we provide the essential framework for calculating the quasinormal-mode spectrum and the relaxation time of a Schwarzschild black hole enveloped in a superconducting plasma. This work thus bridges the gap between condensed-matter physics, particle physics, and gravitational physics, offering a new window into the rich phenomenology of black holes in astrophysical environments~\cite{Bhattacharjee:2019zz,Mahajan:2018zz}.

Quasinormal modes are the complex-frequency damped oscillations that characterize how a perturbed black hole or compact object relaxes to equilibrium, and they provide a key fingerprint in ringdown gravitational-wave signals~\cite{KokkotasSchmidt1999,BertiCardosoStarinets2009}. The Wentzel--Kramers--Brillouin (WKB) approximation gives a semi-analytic route to QNM frequencies by treating the perturbation equations as a quantum-mechanical barrier problem~\cite{SchutzWill1985,IyerWill1987,Konoplya2003,Konoplya2004,Konoplya2011,Konoplya2019,Konoplya2023}, while Prony's method extracts those frequencies directly from time-domain data by fitting a sum of damped exponentials~\cite{Prony1795,BertiCardosoWill2006}. As shown in Ref.~\cite{Toshmatov:2018tyo}, nonlinear effects in the electromagnetic perturbations cause the waveform to be longer-lived. 

The purpose of the present work is to establish a covariant bridge between London electrodynamics and the theory of massive-vector black-hole perturbations. We derive the London--Maxwell equations in curved spacetime, identify their relation to an effective Proca system, and obtain the corresponding covariant wave equation. We then formulate the linear perturbation problem in a general static and spherically symmetric geometry, separate the axial and polar sectors, derive their radial equations and effective potentials, and impose the appropriate quasinormal-mode boundary conditions. Finally, we investigate how the effective London mass modifies the oscillation frequencies, damping rates, polarization structure, and possible quasibound-state spectrum of the electromagnetic field.

The paper is organized as follows. In Sec.~\ref{Sec:1}, we first consider the covariant London and Maxwell equations as a framework for describing electromagnetic wave propagation in a superconducting medium. In Sec.~\ref{Sec:2}, we turn our attention to variable separation and to decoupling the covariant Maxwell--London equations in Schwarzschild spacetime, and we analyze the electromagnetic perturbations in the presence of the superconducting plasma in the axial and polar sectors. Section~\ref{Sec:3} is devoted to the numerical methods, and we report the obtained results in Sec.~\ref{sec:res}. Finally, in Sec.~\ref{Conclusions} we summarize our findings. Throughout the paper, we use the metric signature $(-,+,+,+)$. Greek indices run over spacetime coordinates, $\nabla_{\mu}$ denotes the metric-compatible covariant derivative, and units $c=G=\hbar=1$ are used unless explicitly stated.

\section{Theoretical framework}\label{Sec:1}

The superconducting plasma model around a Schwarzschild black hole represents an important extension of the standard plasma formalism, with potential applications to exotic astrophysical environments such as neutron-star magnetospheres, quark--gluon plasmas, or hypothetical superconducting phases in dense matter near black holes.

\subsection{Covariant London equations}

In four-dimensional curved spacetime, the covariant London equation provides a constitutive relation between the superconducting four-current $J^\mu$ and the electromagnetic four-potential $A^\mu$. It can be written as~\cite{London1935}
\begin{align}\label{London}
J_s^\mu=-\frac{e^2 n_s}{m_s}A^\mu ,
\end{align}
where $e$ and $m_s$ denote, respectively, the charge and effective mass of the superconducting charge carriers, while $n_s(\mathbf{r})$ represents their spatially varying number density. The negative sign reflects the diamagnetic response of the superconducting medium, whereby the induced current opposes the applied electromagnetic field.

\subsection{Covariant Maxwell equations}

The covariant Maxwell equations in curved spacetime read
\begin{align}
\nabla_\nu F^{\nu\mu}=-4\pi J^\mu ,
\end{align}
where $\nabla_\nu$ denotes the covariant derivative associated with the spacetime geometry. The electromagnetic field-strength tensor is defined by $F_{\mu\nu}=\nabla_\mu A_\nu-\nabla_\nu A_\mu$, where the equality follows from the symmetry of the Levi-Civita connection. The total four-current is decomposed as $J^\mu=J_s^\mu+J_e^\mu$, where $J_s^\mu$ represents the superconducting current carried by the condensate, while $J_e^\mu$ denotes the ordinary electric or conduction current associated with non-superconducting charge carriers.

In the present analysis, we assume that the superconducting current constitutes the dominant contribution to the total electric current, such that $|J_s^\mu|\gg|J_e^\mu|$ and/or $J^\mu\simeq J_s^\mu$. Substituting the covariant London relation into Maxwell's equations then yields
\begin{align}\label{ME}
\nabla_\nu F^{\nu\mu}=\omega_s^2 A^\mu ,
\end{align}
where
\begin{align}
\omega_s=\sqrt{\frac{4\pi e^2 n_s}{m_s}}
\end{align}
defines the superconducting plasma (plasmon) frequency. This quantity is related to the London penetration depth $\lambda_s$ through
\begin{align}
\omega_s=\frac{1}{\lambda_s},\qquad\lambda_s=\sqrt{\frac{m_s}{4\pi e^2 n_s}} .
\end{align}
Consequently, the coupled Maxwell--London system takes the form of a Maxwell--Proca equation, with $\omega_s$ acting as an effective mass scale for the electromagnetic field inside the superconducting medium. When the superconducting carrier density is spatially inhomogeneous, both $\omega_s$ and $\lambda_s$ become position-dependent quantities. The constitutive relation~\eqref{London} encodes the covariant manifestation of the Meissner effect in curved spacetime. Within the superconducting region, the electromagnetic field is exponentially screened over the characteristic London penetration depth.

Acting with the covariant derivative $\nabla_\mu$ on both sides of Eq.~\eqref{ME}, one obtains
\begin{align}
\nabla_\mu\nabla_\nu F^{\nu\mu}=\nabla_\mu(\omega_s^2 A^\mu)\equiv0 ,
\end{align}
which allows us to write the following Lorentz-like gauge condition:
\begin{align}\label{Gauge1}
\nabla_\mu A^\mu+A^\mu\nabla_\mu\ln\omega_s^2=0 .
\end{align}
Using the Ricci identity $[\nabla_\nu,\nabla_\mu]A^\nu=R^\nu_{\ \mu}A^\mu$, the left-hand side of Eq.~\eqref{ME} can be simplified as
\begin{align}\nonumber\label{Ricci}
\nabla_\nu F^{\nu\mu}&=\nabla_\nu\nabla^\nu A^\mu-\nabla_\nu\nabla^\mu A^\nu\\
&=\nabla_\nu\nabla^\nu A^\mu-(\nabla^\mu\nabla_\nu A^\nu+R^\nu_{\ \mu}A^\mu) .
\end{align}
Finally, taking into account Eqs.~\eqref{Gauge1} and~\eqref{Ricci}, the Maxwell--London equations for the vector potential can be written as
\begin{align}\label{eqA}
\Box A^\mu-R^\mu_{\ \nu}A^\nu-\omega_s^2A^\mu+\nabla^\mu(A^\nu\nabla_\nu\ln\omega_s^2)=0 ,
\end{align}
where $\Box\equiv\nabla_\mu\nabla^\mu$ is the d'Alembertian in curved spacetime and $R_{\mu\nu}$ is the Ricci tensor. It should be noted that Eq.~\eqref{eqA} reduces to the Proca equation in curved spacetime for a homogeneous superconducting plasma with effective mass $\omega_s=\text{const}$.

\subsection{Plasma distribution}

It should also be emphasized that, in general, the plasma frequency depends on the spatial coordinates $\mathbf{r}=(r,\theta,\phi)$; in that case, solving Eq.~\eqref{eqA} becomes quite difficult. Therefore, for simplicity, we assume that the plasma density depends only on the radial coordinate, i.e.\ $\omega_s=\omega_s(r)$.

For concreteness, we consider the following plasma distributions: (i) homogeneous, (ii) isothermal gas, (iii) exponential, (iv) Gaussian, and (v) power-law. The density of these plasma models can be expressed in the form
\begin{align}\label{Nr}
\frac{n_s(r)}{n_0}=\frac{\omega_s^2(r)}{\omega_0^2}=
\left\{
\begin{array}{lcr}
&1 & \text{Homogeneous} , \\
\\
&\left(1+\frac{r_0^2}{r^2}\right)^{-1} & \text{Isothermal} , \\
\\
&e^{-r/r_0} & \text{Exponential} , \\
\\
&e^{-r^2/r_0^2} & \text{Gaussian} , \\
\\
&\left(\frac{r_0}{r}\right)^h,\quad h\in\mathbb{R} & \text{Power-law} ,
\end{array}
\right.
\end{align}
where $r$ is the radial distance from the central object, and $n_0$, $\omega_0$, and $r_0$ are constants of the plasma models that take different values for each model.

\subsection{Background spacetime}

We now consider electromagnetic perturbations in the Schwarzschild spacetime, analyzing both the axial and polar sectors. The diagonal metric coefficients in the Schwarzschild spacetime are given by
\begin{align}\label{gdd}
g_{\mu\nu}=\text{diag}\left(-f,f^{-1},r^2,r^2\sin^2\theta\right) ,
\end{align}
where $f(r)=1-2M/r$ and $M$ is the black-hole mass. It is worth noting that the Schwarzschild spacetime is a vacuum solution of the Einstein field equations, satisfying $R_{\mu\nu}=0$, which implies that the second term in Eq.~\eqref{eqA} vanishes.

\section{Separation of variables and decoupling}\label{Sec:2}

Here, we aim to investigate how the spatial distribution of the plasmon affects the corresponding perturbation spectra in each of the two sectors. The axial (odd-parity) sector satisfies a single master equation, while the polar (even-parity) sector consists of two coupled equations. The solution for the vector field can be decomposed as~\cite{Rosa:2011PRD}
\begin{align}\label{Sol}
A_\mu=\frac{1}{r}\sum_{i=0}^3\sum_{\ell m}c_i u_{(i)}^{\ell m}(t,r)Z_{\mu}^{(i)\ell m}(r,\theta,\phi) ,
\end{align}
with the following suitable choice of coefficients:
\begin{align}
c_0=c_1=1,\qquad c_2=c_3=\frac{1}{\sqrt{\ell(\ell+1)}} ,
\end{align}
where $u_{(i)}^{\ell m}(t,r)$ are profile functions and $Z_{\mu}^{(i)\ell m}(r,\theta,\phi)$ form a basis of four vector spherical harmonics defined as
\begin{subequations}
\begin{align}
&Z_{\mu}^{(0)\ell m}=\big[1,0,0,0\big]Y_{\ell}^m , \\
&Z_{\mu}^{(1)\ell m}=\big[0,f^{-1},0,0\big]Y_{\ell}^m , \\
&Z_{\mu}^{(2)\ell m}=\frac{r}{\sqrt{\ell(\ell+1)}}\big[0,0,\partial_\theta,\partial_\phi\big]Y_{\ell}^m , \\
&Z_{\mu}^{(3)\ell m}=\frac{r}{\sqrt{\ell(\ell+1)}}\big[0,0,\tfrac{1}{\sin\theta}\partial_\phi,-\sin\theta\,\partial_\theta\big]Y_{\ell}^m .
\end{align}
\end{subequations}
Here, $Y_{\ell}^m(\theta,\phi)$ are the spherical harmonics satisfying
\begin{align}\nonumber
\left[\frac{1}{\sin\theta}\frac{\partial}{\partial\theta}\left(\sin\theta\frac{\partial}{\partial\theta}\right)+\frac{1}{\sin^2\theta}\frac{\partial^2}{\partial\phi^2}+\ell(\ell+1)\right]Y_{\ell}^m=0 ,
\end{align}
with the orthonormality condition $\int Y_{\ell}^{*m}Y_{\ell'}^{m'}\,d\Omega=\delta_{\ell}^{\ell'}\delta_{m}^{m'}$. This property further permits us to cast the orthonormality relation for the functions $Z_{\mu}^{(i)\ell m}(r,\theta,\phi)$ in the form
\begin{align}
\int Z_{\mu}^{*(i')\ell'm'}\eta^{\mu\nu}Z_{\nu}^{(i)\ell m}\,d\Omega=\delta_{i}^{i'}\delta_{\ell}^{\ell'}\delta_{m}^{m'} ,
\end{align}
where
\begin{align}
\eta^{\mu\nu}=\text{diag}\left(1,f^2,\frac{1}{r^2},\frac{1}{r^2\sin^2\theta}\right) .
\end{align}

To see the transformation rules from a slightly different angle: spatial reflection through the origin reverses the orientation of the position vector but leaves the radial distance untouched. While the radial coordinate $r$ stays fixed, the unit vectors spanning the tangent space behave oppositely---the radial unit vector $\hat{r}$ flips sign, whereas the angular unit vectors $\hat{\theta}$ and $\hat{\phi}$ retain their directions under the coordinate mapping $\theta\to\pi-\theta$, $\phi\to\phi+\pi$ (up to the subtle change in the azimuthal basis that introduces the minus sign in $A_\phi$). Consequently, any generic vector harmonic constructed from the gradient of a scalar inherits the simple parity $(-1)^l$, exactly like the spherical harmonics themselves. The fourth harmonic is built from the curl or cross product of the position vector with such a gradient; its axial nature introduces an extra geometric sign flip, shifting its parity to $(-1)^{l+1}$. The even-parity (electric) sector is often associated with ``polar'' disturbances that resemble radial expansions and contractions, where the fluid or field lines lie in the meridional planes. In gravitational physics, these are the modes that couple to density fluctuations and pressure gradients. The odd-parity (magnetic) sector, on the other hand, corresponds to ``axial'' or ``toroidal'' disturbances, which involve twisting, shearing, or rotational motions around the sphere without any radial displacement. They are notoriously harder to excite by simple scalar sources and often dominate in purely transverse wave phenomena, such as certain polarizations of gravitational radiation.

To derive the master equations for the electromagnetic perturbations, we must first eliminate the gauge freedom inherent in the Maxwell--London field equations. For the polar sector of the electromagnetic perturbations, we impose the following Lorentz-like gauge condition:
\begin{align}\label{Gauge2}
-\dot u_{(0)}+u_{(1)}'+\frac{f}{r}\left(u_{(1)}-u_{(2)}\right)+\frac{2\omega_s'(r)}{\omega_s(r)}u_{(1)}=0 ,
\end{align}
which is valid for the polar sector. Now, by inserting the solution~\eqref{Sol} into the main equation~\eqref{eqA} and using the gauge condition~\eqref{Gauge2}, the set of equations for the functions $u_{(i)}(t,r)$ can be expressed as
\begin{subequations}\label{main}
\begin{align}
\mathcal{D}u_{(0)}&+f'(\dot u_{(1)}-u_{(0)}')+\frac{2\omega_s'}{\omega_s}\dot u_{(1)}=0 , \\
\mathcal{D}u_{(1)}&+\frac{2\omega_s'}{\omega_s}u_{(1)}'-\frac{2f}{r^2}\left(1-\frac{3M}{r}\right)(u_{(1)}-u_{(2)})\nonumber\\
&+2\left(\frac{\omega_s''}{\omega_s}-\frac{\omega_s'^2}{\omega_s^2}-\frac{f}{r}\frac{\omega_s'}{\omega_s}\right)u_{(1)}=0 , \\
\mathcal{D}u_{(2)}&+\frac{2\ell(\ell+1)}{r^2}\left(f+\frac{r\omega_s'}{\omega_s}\right)u_{(1)}=0 , \\
\mathcal{D}u_{(3)}&=0 ,
\end{align}
\end{subequations}
where a dot denotes the time derivative and a prime denotes the derivative with respect to the tortoise coordinate. The differential operator $\mathcal{D}$ is defined as
\begin{align}
\mathcal{D}=-\frac{\partial^2}{\partial t^2}+\frac{\partial^2}{\partial r_*^2}-f\left[\frac{\ell(\ell+1)}{r^2}+\omega_s^2(r)\right] ,
\end{align}
where $r_*$ is the tortoise coordinate defined as $r_*=\int dr/f$.

\begin{figure}
\centering
\includegraphics[width=\hsize]{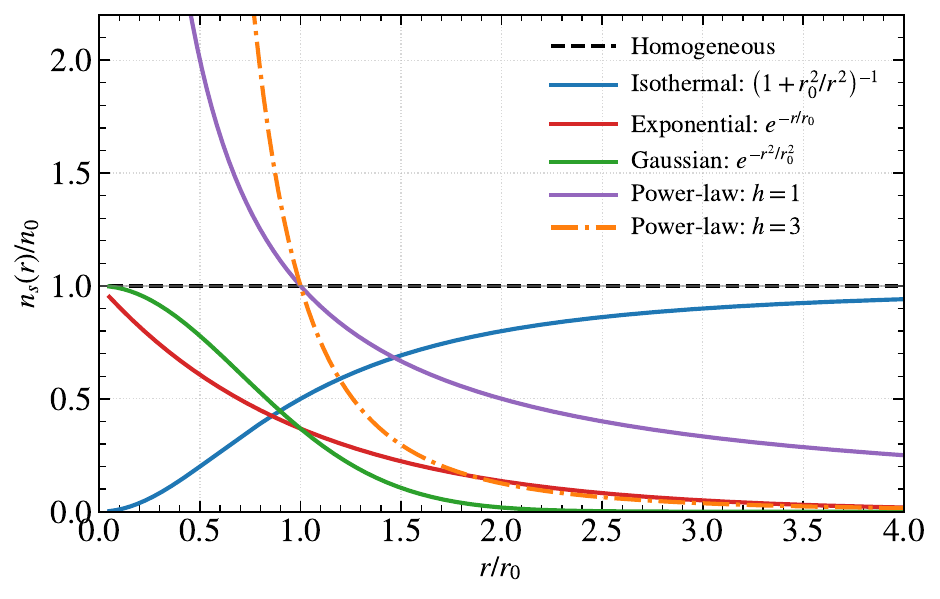}
\includegraphics[width=\hsize]{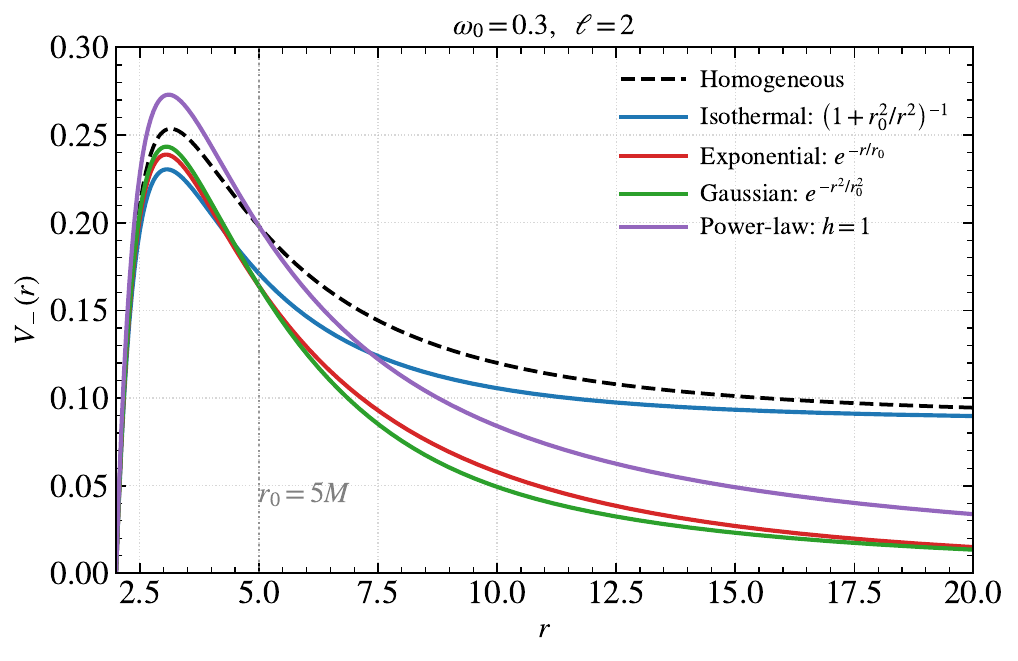}
\caption{(Top panel) Radial dependence of the different types of superconducting plasma distribution. (Bottom panel) Radial dependence of the effective potential $V_-(r)$ for axial vector perturbations for different superconducting plasma distributions.}\label{fig1}
\end{figure}

\subsection{Axial sector}

In previous analyses, we demonstrated that the axial perturbations involve only the angular components of the vector potential and that the equation governing the profile function is fully decoupled. Accordingly, the Schr\"odinger-like equation for the mode $\psi_-(t,r)=u_{(3)}(t,r)$ in Schwarzschild spacetime in the presence of a superconducting plasma environment can be written as
\begin{align}\label{AxWave}
\left[\frac{\partial^2}{\partial t^2}-\frac{\partial^2}{\partial r_*^2}+V_-(r)\right]\psi_-(t,r)=0 ,
\end{align}
where $V_-(r)$ is the effective potential for axial electromagnetic perturbations, defined as
\begin{align}\label{Vax}
V_-(r)=f(r)\left[\frac{\ell(\ell+1)}{r^2}+\omega_s^2(r)\right] .
\end{align}
Taking into account the plasma distribution given in Eq.~\eqref{Nr}, it is clear that the asymptotic behavior of the effective potential above is governed by the superconducting plasma frequency, which gives
\begin{align}\label{Vasymp}
\lim_{r\to\infty}V_-(r)=
\left\{
\begin{array}{lcr}
&\omega_0^2 , & \text{Homogeneous} , \\
\\
&\omega_0^2 , & \text{Isothermal} , \\
\\
&0 , & \text{Exponential} , \\
\\
&0 , & \text{Gaussian} , \\
\\
&0 , & \text{Power-law} .
\end{array}
\right.
\end{align}

\begin{figure*}
\centering
\includegraphics[width=\hsize]{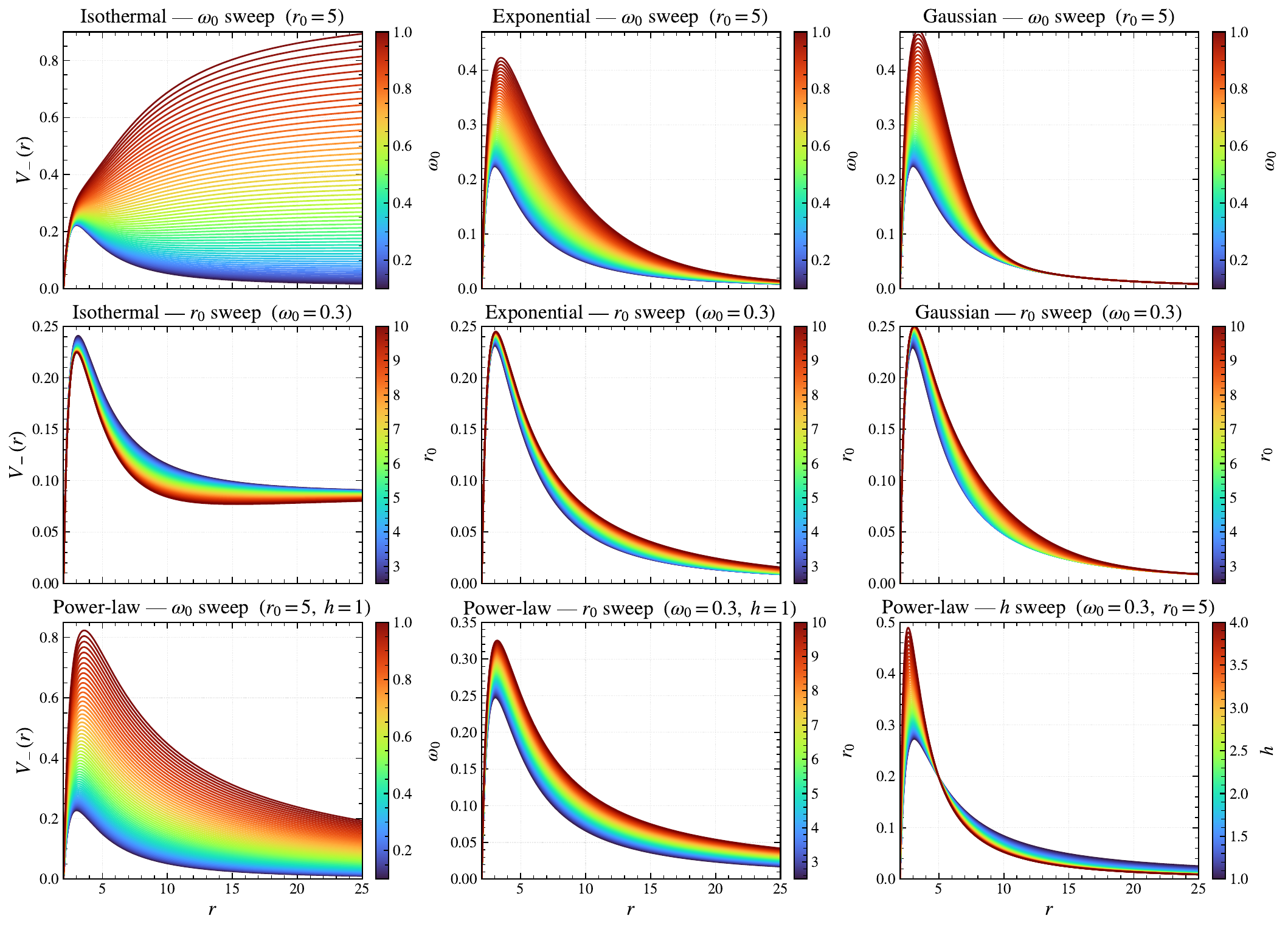}
\caption{Colour-coded families of the axial effective potential $V_{-}(r)$ for electromagnetic perturbations in a Schwarzschild background surrounded by plasma. Each panel displays the potential as a function of the areal radius $r$, with the colour encoding the value of the parameter being varied; the horizon is marked by the grey dashed line at $r_{h}=2M$. The top row shows the dependence on the plasma frequency $\omega_{0}$ at fixed scale length $r_{0}=5M$ for the isothermal (left), exponential (middle), and Gaussian (right) distributions. The middle row shows the dependence on the plasma scale length $r_{0}$ at fixed $\omega_{0}=0.3M^{-1}$ for the same three profiles. The bottom row corresponds to the power-law distribution $\omega_{s}^{2}\propto(r_{0}/r)^{h}$: the left and middle panels show the variation with $\omega_{0}$ (at fixed $r_{0}=5M$, $h=1$) and with $r_{0}$ (at fixed $\omega_{0}=0.3M^{-1}$, $h=1$), respectively, while the right panel shows the dependence on the power-law index $h$ at fixed $r_{0}=5M$ and $\omega_{0}=0.3M^{-1}$. Throughout, the multipole number is $\ell=2$.}\label{fig2}
\end{figure*}

In Fig.~\ref{fig1}, we illustrate the radial dependence of the superconducting plasma distribution and the corresponding effective potential for axial electromagnetic perturbations in the presence of the superconducting plasma environment for different values of the parameters. For simplicity, we assume that $r_0$ is the same in all cases. The graph shows that the effective potential vanishes at the Schwarzschild radius ($r=2M$), rises to its maximum value, and then decreases as the radial distance becomes large. For homogeneous and isothermal-gas plasma distributions, the effective potential tends to $\omega_0^2$, while for all other cases it vanishes, as confirmed by Eq.~\eqref{Vasymp}.

In Fig.~\ref{fig2}, we plot the effective potential as a function of the radial coordinate for axial electromagnetic perturbations in the presence of isothermal, exponential, and Gaussian-type superconducting plasma environments for different values of the parameters. It is evident that, as the plasma frequency increases, the peak of the effective potential increases in all cases considered. This indicates that a larger plasma frequency generally leads to a stronger effective-potential structure, which is favourable for supporting a localized mode. However, for a particular value of the plasma frequency $\omega_s$, the effective potential loses its peak in the homogeneous and isothermal-gas models. Once this occurs, the effective potential no longer provides the spatial variation required to trap a mode; instead, it becomes constant or monotonic. Consequently, Eq.~\eqref{AxWave} no longer admits an eigenvalue solution. In other words, the discrete eigenvalue disappears, and the corresponding trapped eigenmode cannot exist. Physically, this transition can be viewed as a cutoff for the existence of the eigenmode: the mode is supported only when the effective potential retains the required peak. Thus, the existence of the eigenvalue is restricted to those parameter regimes in which the effective potential still possesses a sufficiently pronounced peak.

It is important to note that the effective potential for electromagnetic perturbations in the Schwarzschild spacetime contains a stationary point at $r_0=3M$; however, our analysis shows that this stationary point depends on the plasma distribution. In the case of a homogeneous plasma, one finds that the stationary point of the effective potential~\eqref{Vax} is located at
\begin{align}
r_0=\frac{\ell(\ell+1)}{2M\omega_0^2}\left(1-\sqrt{1-\frac{12M^2\omega_0^2}{\ell(\ell+1)}}\right) ,
\end{align}
which reduces to $r_0=3M$ when $\omega_0$ vanishes. Our analysis shows that the maximum value of the stationary point is located at $r_0=6M$, which means $3M\leq r_0\leq 6M$ for any values of $\omega_0$ and $\ell$. It should be emphasized that for the other types of plasma distribution, the stationary points are determined by the condition $V'(r)=0$.

\subsection{Polar sector}

We now focus on the vector perturbations in the polar sector. In this case, the derivation of a Schr\"odinger-like equation for a single master function is quite difficult; however, for a particular case, the problem can be solved simply. Here, the electromagnetic perturbations are governed by a system of coupled equations involving the profile functions $u_{(0)}$, $u_{(1)}$, and $u_{(2)}$. Unlike the axial sector, which admits a single decoupled master equation, the polar sector requires careful treatment due to the coupling between the different components of the vector potential. This coupling arises from the broken gauge invariance when a nonzero mass or effective mass is present, introducing additional longitudinal degrees of freedom. These modes are characterized by radial and temporal components of the vector potential, as well as angular gradient components, and are associated with ``electric''-type perturbations that couple to density fluctuations and pressure gradients in the surrounding medium. It is evident from Eqs.~\eqref{main} that the polar sector does not reduce to a single master equation. Instead, the three profile functions $u_{(0)}$, $u_{(1)}$, and $u_{(2)}$ are dynamically coupled through terms proportional to $1/r^2$ and $f^2/r^2$.

In the vacuum limit, $\omega_s\to0$, the vector field becomes an ordinary electromagnetic field, and gauge freedom is recovered. In this case, the polar sector separates into two independent families of solutions: the ``vector'' modes, which are physical and describe transverse degrees of freedom, and the ``scalar'' modes, which are purely gauge and therefore do not correspond to physical radiation. To demonstrate this separation, one introduces the variable
\begin{align}
\psi=u_{(2)}'-\frac{\ell(\ell+1)}{r}u_{(1)} ,
\end{align}
and substitutes it into the polar perturbation equations. In the vacuum limit, this gives a decoupled wave equation for $\psi$, namely $\mathcal{D}\psi=0$. This equation describes the physical vector modes of the electromagnetic field and has the same form as the equation governing the axial sector in the vacuum case. In the presence of plasma, the polar sector exhibits a rich and nontrivial structure. The coupling between the ``vector'' and ``scalar'' solutions becomes dynamically important, leading to a mixing that depends on the plasma frequency $M\omega_s$.

Depending on the specific gauge choice and the nature of the $\omega_s$ dependence, the polar sector is governed by a coupled system of three equations, which can be reduced to a single wave equation with an effective potential $V_+(r)$ after imposing the appropriate gauge condition and eliminating the redundant degrees of freedom. In the monopole mode ($\ell=0$), the Schr\"odinger-like equation in the polar sector can be derived as
\begin{align}\label{PolWave}
\left[\frac{\partial^2}{\partial t^2}-\frac{\partial^2}{\partial r_*^2}+V_+(r)\right]\psi_+(t,r)=0 ,
\end{align}
where $\psi_+(t,r)=\omega_s u_{(1)}(t,r)$ is a new radial profile function and $V_+(r)$ is the effective potential in the polar sector in the monopole case, defined as
\begin{align}\label{eqpotpolar}
V_+(r)=f\left[\frac{2}{r^2}\left(1-\frac{3M}{r}\right)+\frac{2\omega_s'}{r\omega_s}+\omega_s^2\right]+\frac{2\omega_s'^2}{\omega_s^2}-\frac{\omega_s''}{\omega_s} .
\end{align}

\begin{figure}
\centering
\includegraphics[width=\hsize]{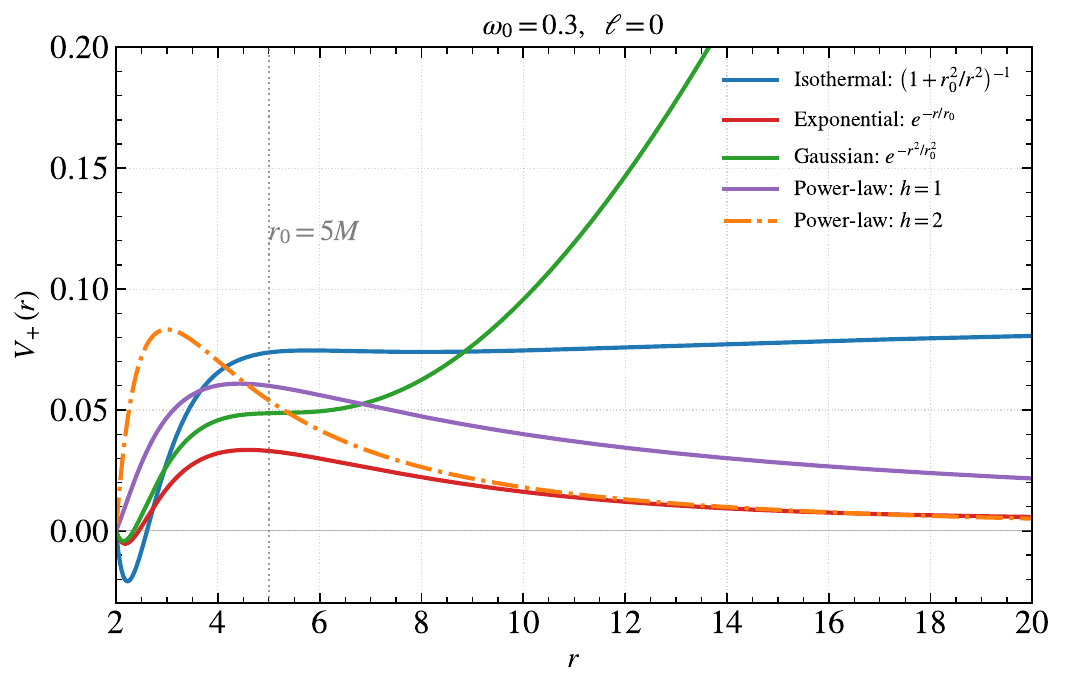}
\caption{Radial dependence of the effective potential $V_-(r)$ for axial vector perturbations for different superconducting plasma distributions.}\label{fig3}
\end{figure}

Figure~\ref{fig3} displays the polar effective potential $V_{+}(r)$ for the four plasma distributions considered in this work, illustrating how the choice of profile affects the shape of the barrier. For $\omega_{0}=0.3M^{-1}$, the isothermal, exponential, and Gaussian models all develop a negative region in the immediate vicinity of the horizon. Such a negative dip signals an instability of the corresponding polar modes, since the effective potential is no longer bounded from below in the near-horizon region.

The asymptotic behaviour of the tails distinguishes the four profiles more sharply. The isothermal potential approaches a constant value asymptotically, reflecting the slow, power-law decay of $\omega_{s}^{2}(r)$ at large $r$. The exponential potential reproduces the familiar form of the standard Schwarzschild barrier, with a single positive maximum and a rapid, exponential fall-off in both directions; in this sense, it is the closest analogue of the well-known vacuum case. The Gaussian potential behaves differently: beyond $r\simeq8M$ it rises steeply, so that the effective barrier is no longer localized and the notion of a single dominant peak breaks down. The power-law profile, by contrast, exhibits no such pathologies---the potential remains positive, single-peaked, and well localized over the entire range of $r$ considered here, making it the most well-behaved of the four models in the polar sector.

\begin{figure*}
\centering
\includegraphics[width=\hsize]{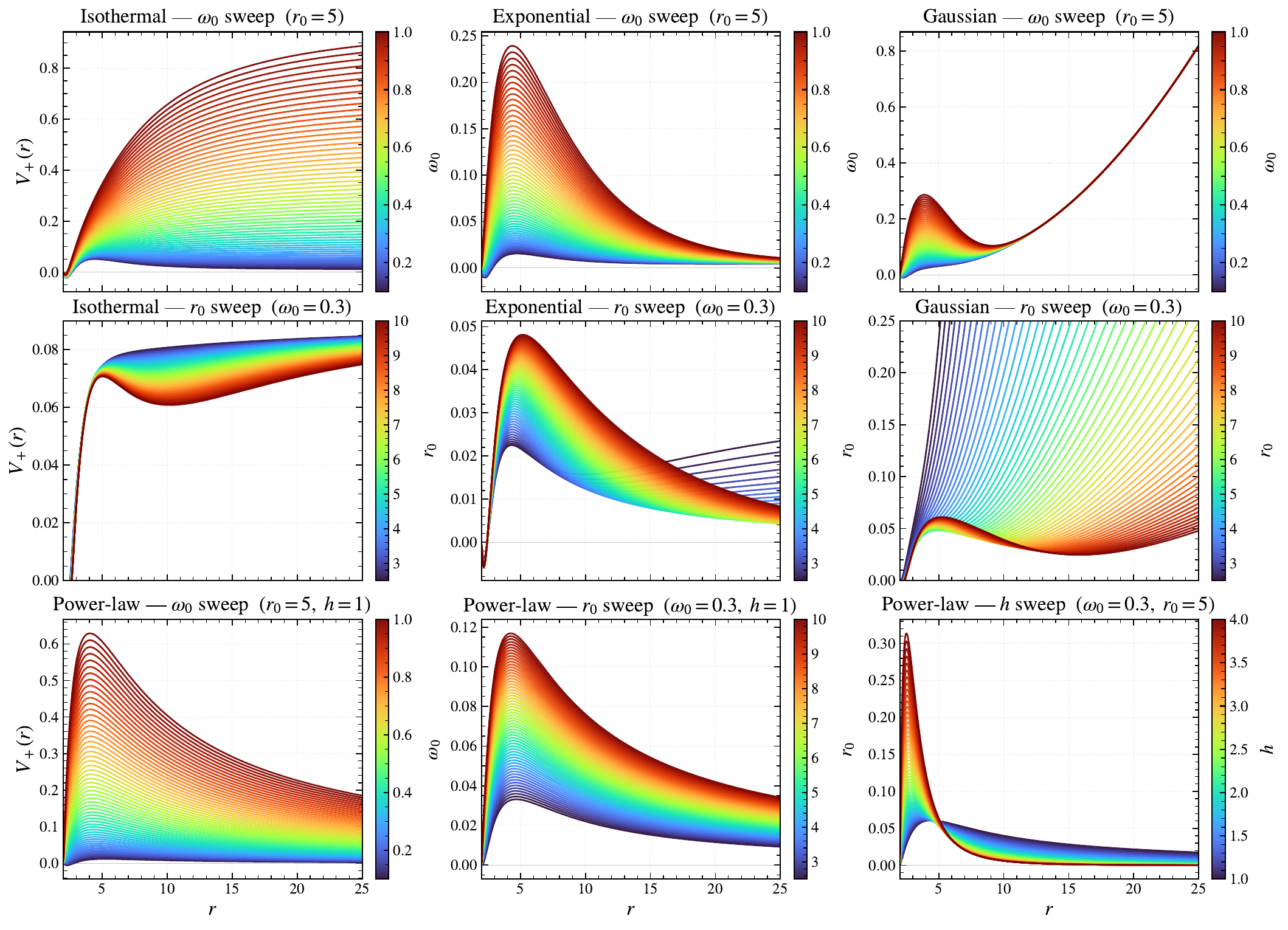}
\caption{Colour-coded families of the polar effective potential $V_{+}(r)$ for electromagnetic perturbations in a Schwarzschild background surrounded by plasma. Panels are arranged as in Fig.~\ref{fig2}, with the top row varying $\omega_{0}$, the middle row varying $r_{0}$, and the bottom row showing the power-law distribution with $\omega_{0}$ (left), $r_{0}$ (middle), and the index $h$ (right) swept. The horizon is marked by the grey dashed line at $r_{h}=2M$, and $\ell=2$ throughout. In contrast to the axial case, the derivative terms in $V_{+}$ can produce a steep rise or a negative dip near the horizon.}\label{fig4}
\end{figure*}

The behaviour of the polar potential is illustrated in Fig.~\ref{fig4}, where we show $V_{+}(r)$ for the same isothermal, exponential, Gaussian, and power-law plasma distributions considered in the axial sector, with the same parameter ranges for $\omega_{0}$, $r_{0}$, and $h$.

Far from the horizon, the polar and axial potentials coincide. This is expected: at large $r$, the plasma enters the wave equation only through the local mass term $\omega_{s}^{2}(r)$, which is insensitive to the parity of the perturbation, and the derivative corrections in Eq.~\eqref{eqpotpolar} fall off faster than $\omega_{s}^{2}$ itself. The asymptotic tails of the two potentials are therefore identical for every profile, and the differences between them are confined to a region of a few horizon radii around $r_{h}$.

Inside this region, the two sectors diverge. While $V_{-}$ retains a single positive barrier in all cases, $V_{+}$ acquires a shape that depends sensitively on the profile and on the values of $r_{0}$ and $\omega_{0}$. Three trends are visible in Fig.~\ref{fig4}. First, increasing $\omega_{0}$ at fixed $r_{0}$ raises the overall amplitude of the potential, in agreement with the $\omega_{s}^{2}$ and $\omega_{s}'^{2}/\omega_{s}^{2}$ contributions that scale quadratically with the plasma frequency. Second, increasing $r_{0}$ at fixed $\omega_{0}$ broadens the potential and pushes its maximum outward, exactly as in the axial sector. Third, and most importantly, the near-horizon behaviour depends on how rapidly $\omega_{s}$ varies with $r$: for profiles whose logarithmic derivative $\omega_{s}'/\omega_{s}$ grows slowly, the potential remains positive and rises monotonically toward $r_{h}$, whereas for steeper profiles the curvature term $\propto(f\omega_{s}')'/\omega_{s}$ becomes dominant and $V_{+}$ develops a negative dip immediately outside the horizon.

The power-law family makes this last point particularly clear. As the index $h$ increases from $1$ to $4$ at fixed $\omega_{0}$ and $r_{0}$, the potential becomes increasingly concentrated near the horizon, and its near-horizon behaviour changes from a monotonic rise to a sharp negative excursion. The exponential and Gaussian profiles exhibit the same qualitative transition, but at smaller values of $\omega_{0}$ and $r_{0}$, because their logarithmic derivatives are milder than those of the power law at small $r$.

\section{Numerical methods}\label{Sec:3}

In this section, we study numerical results for the quasinormal modes of electromagnetic perturbations in Schwarzschild spacetime for different configurations of the superconducting plasma environment. As mentioned before, the superconducting plasma distribution falls off with distance for the Gaussian, exponential, and power-law models, $\lim_{r\to\infty}\omega_s=0$, while for the uniform and isothermal-gas models the plasma distribution is constant, $\lim_{r\to\infty}\omega_s=\omega_0$. To solve the Schr\"odinger-type wave equations~\eqref{AxWave} and~\eqref{PolWave}, we adopt the following boundary constraints: the wave must exhibit purely ingoing behaviour at the event horizon and purely outgoing behaviour at spatial infinity:
\begin{align}
&\psi_\pm\sim e^{-i\omega r_{*}} , \qquad r_*\to-\infty , \\
&\psi_\pm\sim e^{i\omega r_{*}} , \qquad r_*\to\infty .
\end{align}

\subsection{WKB method}

One of the simplest semi-analytical approaches for producing a solution to the Schr\"odinger-like equation is the Wentzel--Kramers--Brillouin (WKB) method. For frequency-domain analysis, we apply the semi-analytical sixth-order WKB method with Pad\'e approximation, following the same strategy detailed in the previous section. The higher-order WKB formula~\cite{PhysRevD.96.024011}
\begin{align}
\omega^2 ={}& V_0 + A_2(K^2) + A_4(K^2) + A_6(K^2) \nonumber\\
&- iK\bigl(-2V_1 + A_3(K^2) + A_5(K^2) + A_7(K^2)\bigr) + \dots
\end{align}
($K=n+1/2$, $n=0,1,2,\dots$) is used, and the series is improved by forming the Pad\'e approximant $\tilde{P}_{\tilde{n}/\tilde{m}}(\epsilon)$ with $\tilde{n}+\tilde{m}=6$. For the fundamental mode ($n=0$), near-diagonal approximants ($\tilde{n}\approx\tilde{m}$) are chosen, as they yield the best accuracy~\cite{PhysRevD.96.024011}.

\subsection{Time-domain integration}

To validate the WKB results, we also solve the wave equation directly in the time domain. Without assuming a harmonic time dependence, the master equation reduces to
\begin{align}
\frac{\partial^2\psi}{\partial t^2}-\frac{\partial^2\psi}{\partial r_*^2}+V(t,r_*)\psi=0 .
\end{align}
Introducing light-cone coordinates $du=dt-dr_*$, $dv=dt+dr_*$, one obtains
\begin{align}
\bigl(4\,\partial_u\partial_v+V(u,v)\bigr)\psi(u,v)=0 ,
\end{align}
which we discretize on a uniform null grid~\cite{Gundlach:1993tp}:
\begin{align}
\psi(N)={}& \psi(W) + \psi(E) - \psi(S)\nonumber\\
&-\frac{\Delta^2}{8}V(S)\left(\psi(W)+\psi(E)\right)+\mathcal{O}(\Delta^4) .
\end{align}
The initial data are a Gaussian pulse $\psi(u_0,v)=A\exp[-(v-v_c)^2/\sigma^2]$. The dominant quasinormal frequency is then extracted from the ringdown signal using the Prony method~\cite{Prony1795}. This approach has been employed in numerous works and is described in detail in the literature~\cite{Lin:2024ubg,Bolokhov:2023dxq,Cuyubamba:2016cug}.

\section{Results}\label{sec:res}

\subsection{Results for the axial sector}

\begin{figure}
\centering
\includegraphics[width=\hsize]{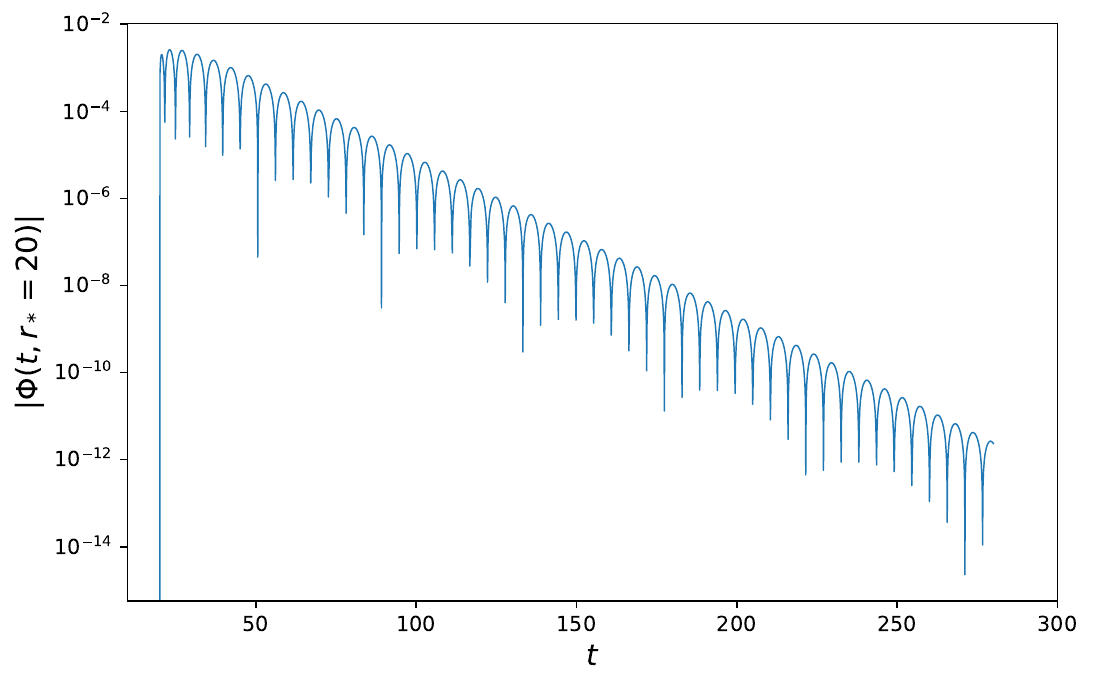}
\caption{Time-domain wave function for a power-law plasma ($\omega_0^2\times r_0=1$, $h=1$), showing a clear ringdown phase followed by a power-law tail.}\label{fig:td_nonuniform}
\end{figure}

As an example, a typical time-domain profile for the power-law plasma case ($\omega_0^2\times r_0=1$, $h=1$) is shown in Fig.~\ref{fig:td_nonuniform}. After the initial burst, the signal enters a well-defined ringdown phase, from which the complex frequency can be extracted accurately.

\begin{figure*}
\centering
\includegraphics[width=0.45\linewidth]{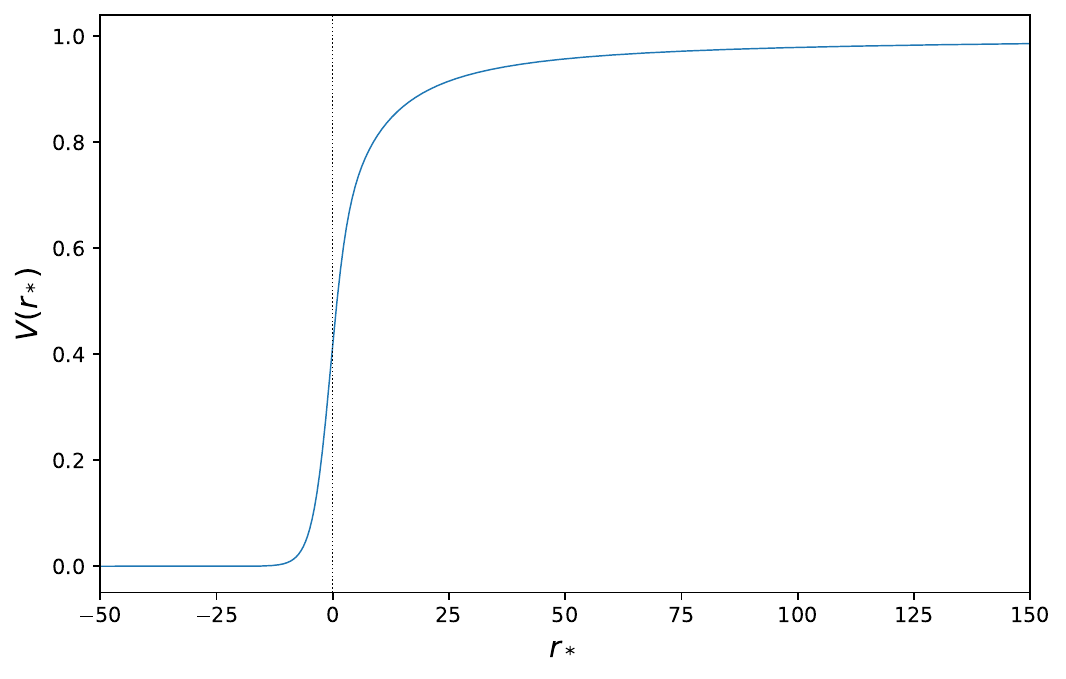}
\includegraphics[width=0.45\linewidth]{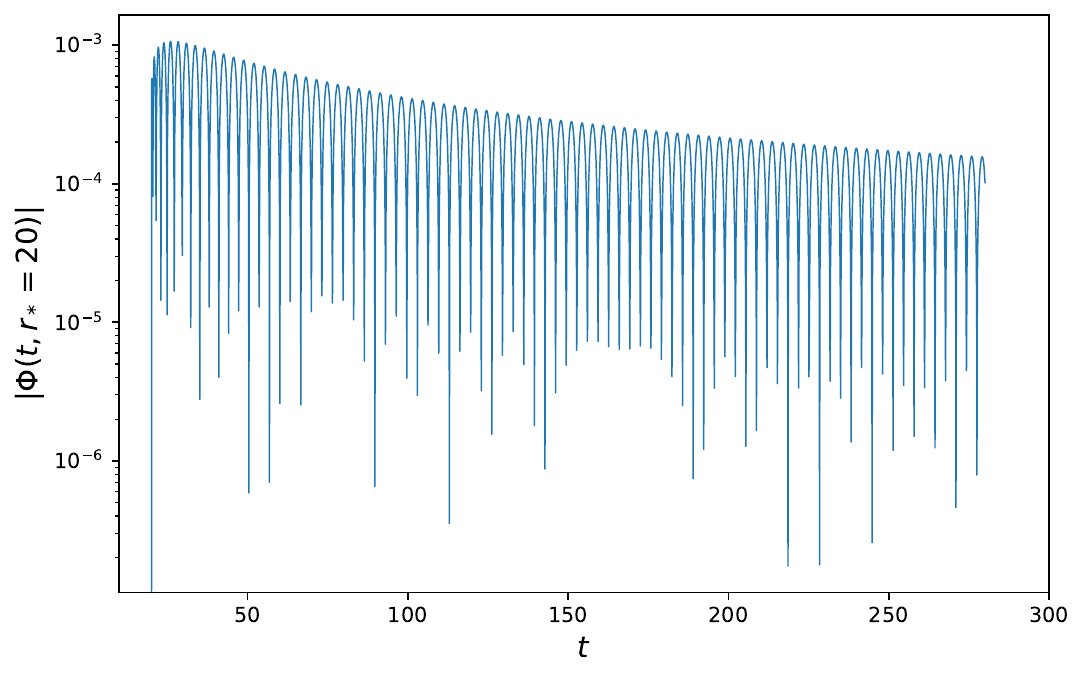}
\caption{Left: scattering potential in tortoise coordinate for a homogeneous plasma ($\omega_0=1$). Right: corresponding time-domain wave function; the extremely slow decay indicates a quasi-resonance.}\label{fig:longlived}
\end{figure*}

The homogeneous plasma is special. Because $\omega_s=\omega_0$ is constant, the effective potential acquires a constant asymptotic value, giving it a step-like shape far away (Fig.~\ref{fig:longlived}, left). This configuration supports very long-lived modes---quasi-resonances---whose imaginary part is orders of magnitude smaller than in the other model cases. The time-domain evolution (Fig.~\ref{fig:longlived}, right) reveals an extremely slow decay; the signal appears almost non-decaying over the plotted interval, making the extraction of the damping rate from the time profile challenging.

\begin{table}[]
\centering
\caption{Eigenvalues ($\ell=2$) obtained by the WKB method for the homogeneous plasma distribution with different values of the model parameter $\omega_0$.}
\begin{ruledtabular}
\begin{tabular}{|c|c|}
$\omega_0^2$ & Homogeneous plasma \\
\hline
$0$ & $0.457595-0.095005 i$ \\
$0.1$ & $0.461045-0.0937849 i$ \\
$0.2$ & $0.471444-0.0900909 i$ \\
$0.3$ & $0.488948-0.0838072 i$ \\
$0.4$ & $0.513836-0.0746965 i$ \\
$0.5$ & $0.546501-0.0623003 i$ \\
\end{tabular}
\end{ruledtabular}
\label{tab:homo1}
\end{table}

\begin{table}
\centering
\caption{WKB calculation ($\ell=2$) for the isothermal-gas model for different values of the parameters $\omega_0$ and $r_0$.}
\begin{ruledtabular}
\begin{tabular}{|c|c|c|}
$r_0$ & $\omega_0$ & Isothermal-gas model \\
\hline
& $0.1$ & $0.458066-0.0944594 i$ \\
& $0.2$ & $0.45945-0.0928013 i$ \\
$6$ & $0.3$ & $0.461643-0.0899837 i$ \\
& $0.4$ & $0.464525-0.0860157 i$ \\
& $0.5$ & $0.467875-0.0810479 i$ \\ \hline
& $0.1$ & $0.457849-0.0946694 i$ \\
& $0.2$ & $0.4586-0.0936628 i$ \\
$8$ & $0.3$ & $0.459816-0.0919859 i$ \\
& $0.4$ & $0.461482-0.0896548 i$ \\
& $0.5$ & $0.463484-0.0866577 i$ \\ \hline
& $0.1$ & $0.457754-0.0947838 i$ \\
& $0.2$ & $0.458228-0.0941219 i$ \\
$10$ & $0.3$ & $0.459008-0.0930195 i$ \\
& $0.4$ & $0.460081-0.0914834 i$ \\
& $0.5$ & $0.461407-0.0895003 i$ \\
\end{tabular}\label{tab:iso1}
\end{ruledtabular}
\end{table}

\begin{table}
\centering
\caption{WKB calculation ($\ell=2$) for the exponential distribution for different values of the parameters $\omega_0$ and $r_0$.}
\begin{ruledtabular}
\begin{tabular}{|c|c|c|}
$r_0$ & $\omega_0$ & Exponential distribution \\
\hline
& $0.1$ & $0.459831-0.0945326 i$ \\
& $0.2$ & $0.466535-0.0931488 i$ \\
$6$ & $0.3$ & $0.477685-0.0909518 i$ \\
& $0.4$ & $0.493229-0.0881101 i$ \\
& $0.5$ & $0.51305-0.0848573 i$ \\ \hline
& $0.1$ & $0.460097-0.0943924 i$ \\
& $0.2$ & $0.46761-0.0925862 i$ \\
$8$ & $0.3$ & $0.480152-0.0896832 i$ \\
& $0.4$ & $0.497733-0.085865 i$ \\
& $0.5$ & $0.520319-0.0814181 i$ \\ \hline
& $0.1$ & $0.460268-0.0942949 i$ \\
& $0.2$ & $0.468304-0.0921918 i$ \\
$10$ & $0.3$ & $0.481746-0.0887804 i$ \\
& $0.4$ & $0.500664-0.0842304 i$ \\
& $0.5$ & $0.525113-0.0788416 i$ \\
\end{tabular}\label{tab:exp1}
\end{ruledtabular}
\end{table}

\begin{table}
\centering
\caption{WKB calculation ($\ell=2$) for the Gaussian distribution for different values of the parameters $\omega_0$ and $r_0$.}
\begin{ruledtabular}
\begin{tabular}{|c|c|c|}
$r_0$ & $\omega_0$ & Gaussian distribution \\
\hline
& $0.1$ & $0.4606-0.0944025 i$ \\
& $0.2$ & $0.469615-0.0927089 i$ \\
$6$ & $0.3$ & $0.484588-0.0902597 i$ \\
& $0.4$ & $0.505501-0.0875504 i$ \\
& $0.5$ & $0.53133-0.0851807 i$ \\ \hline
& $0.1$ & $0.460801-0.0941379 i$ \\
& $0.2$ & $0.470478-0.0916103 i$ \\
$8$ & $0.3$ & $0.486791-0.0875898 i$ \\
& $0.4$ & $0.509796-0.0837682 i$ \\
& $0.5$ & $0.540042-0.0777287 i$ \\ \hline
& $0.1$ & $0.460889-0.0940134 i$ \\
& $0.2$ & $0.47084-0.0910545 i$ \\
$10$ & $0.3$ & $0.487679-0.0862197 i$ \\
& $0.4$ & $0.511884-0.0797165 i$ \\
& $0.5$ & $0.543846-0.0726412 i$ \\
\end{tabular}\label{tab:gaus1}
\end{ruledtabular}
\end{table}

Tables~\ref{tab:homo1}, \ref{tab:iso1}, \ref{tab:exp1}, and~\ref{tab:gaus1} summarize the dependence of the eigenvalues for axial perturbations on $\omega_0$ and $r_0$ for the homogeneous plasma, isothermal gas, exponential, and Gaussian distributions, respectively. For the homogeneous plasma (Table~\ref{tab:homo1}) with $\ell=2$ fixed, increasing $\omega_0$ increases both the real and imaginary parts of the eigenvalues, implying that the modes become unstable. For the isothermal gas (Table~\ref{tab:iso1}), increasing $\omega_0$ likewise raises both parts, while increasing $r_0$ lowers both, thus making the modes more stable. For the exponential distribution (Table~\ref{tab:exp1}), $\omega_0$ follows the same trend as above, but $r_0$ has an effect analogous to that of $\omega_0$: it increases both parts and therefore reduces stability. The Gaussian distribution (Table~\ref{tab:gaus1}) exhibits the same behaviour as the exponential case. Throughout, we adopt the convention $\omega=\omega_R+i\omega_I$, so that an increase in the imaginary part corresponds to a less stable mode.

\begin{table*}[ht]
\centering
\caption{Axial sector: eigenvalues for selected values of the parameters for the power-law model ($\ell=2$).}
\label{table:power1}
\begin{ruledtabular}
\begin{tabular}{c|c c c|c c c}
& & $\omega_0^2\times r_0^h=1$, $h$ changes & & & $h=1$, $\omega_0^2\times r_0$ changes & \\ [0.6ex] \hline
& WKB & Prony & $|\Delta(\%)|$ & WKB & Prony & $|\Delta(\%)|$ \\[0.6ex] \hline
$1$ & $0.569889-0.083636 i$ & $0.569892-0.083624 i$ & $0.0003-0.0012i$ & $0.569889-0.083636 i$ & $0.569892-0.083624i$ & $0.0003-0.0012i$ \\
$2$ & $0.496389-0.095180 i$ & $0.496392-0.095170i$ & $0.0003-0.001i$ & $0.666880-0.077691 i$ & $0.666889-0.077664 i$ & $0.0009-0.0027$ \\
$3$ & $0.470742-0.095908 i$ & $0.470746-0.095902 i$ & $0.0004-0.0006i$ & $0.752785-0.074206 i$ & $0.752804-0.074162 i$ & $0.0019-0.0044i$ \\
$4$ & $0.462020-0.095323 i$ & $0.462023-0.095318i$ & $0.0003-0.0005i$ & $0.830457-0.071964 i$ & $0.830490-0.071910 i$ & $0.0033-0.0054i$ \\[1ex]
\end{tabular}
\end{ruledtabular}
\end{table*}

Table~\ref{table:power1} shows the fundamental quasinormal frequencies ($n=0$) for vector perturbations, computed using the sixth-order WKB method and integration of the time-domain followed by a Prony fit. Plasma parameters are varied as follows: the left half of the table fixes $\omega_0^2\times r_0^h=1$ and varies the fall-off exponent $h$, while the right half maintains $h=1$ and varies $\omega_0^2\times r_0^h$. The last column of each half lists the absolute difference $|\Delta(\%)|$ between the two methods. The agreement is excellent: the differences are typically of order $10^{-3}\%$, confirming the reliability of both techniques. Increasing $h$ decreases both the real and imaginary parts, while increasing $\omega_0^2\times r_0^h$ increases both parts.

\subsection{Results for the polar sector}

The polar potential $V_+(r)$ given in Eq.~\eqref{eqpotpolar} does not admit the same spectral treatment that we have applied to the axial sector. In the vacuum limit $\omega_s\to0$, both potentials reduce to the standard Regge--Wheeler and Zerilli forms and share the same quasinormal spectrum, as isospectrality requires. Once the plasma is present, however, this degeneracy is lifted, and the eigenvalue problem associated with $V_+$ acquires features that preclude a direct application of the WKB and Prony methods employed above.

The asymptotic behaviour of the two potentials already differs for two of the four profiles. Because the derivative terms in Eq.~\eqref{eqpotpolar} involve $\omega_s'/\omega_s$ and $(f\omega_s')'/\omega_s$, the large-$r$ limit of $V_+$ is controlled not only by the local mass term $\omega_s^{2}(r)$ but also by the logarithmic derivative of the plasma frequency. For the isothermal and power-law distributions, $\omega_s'/\omega_s$ decays faster than $1/r$, so $V_+$ approaches zero at spatial infinity, in agreement with $V_-$. For the exponential profile, $\omega_s'/\omega_s$ tends to a nonzero constant, and $V_+$ asymptotes to the finite value $1/(4r_{0}^{2})$, which coincides with the tail of the axial potential only in the limit $r_{0}\to\infty$. For the Gaussian profile, $\omega_s'/\omega_s$ grows linearly in $r$, and $V_+$ diverges at large $r$---a behaviour that has no counterpart in the axial sector and signals the breakdown of any localized-barrier picture for that profile.

Close to the horizon, the situation is equally delicate. Depending on the profile and on the values of $r_{0}$ and $\omega_0$, $V_+$ may either rise steeply as $r\to2M$ or develop a negative region immediately outside the horizon. Either feature violates the single-barrier assumption on which the WKB approximation rests, and the associated prompt, non-exponential response contaminates Prony-type fits of the ringdown. We therefore refrain from extracting polar quasinormal frequencies in the present work and restrict the numerical analysis of the spectrum to the axial sector, where the effective potential retains the single-barrier structure required by both methods and where the plasma corrections to the real and imaginary parts of the eigenvalues can be determined unambiguously.

The polar sector nonetheless admits a well-posed eigenvalue problem when formulated in terms of quasibound states, that is, solutions of the radial equation that are regular at the horizon and decay at spatial infinity. For plasma profiles whose near-horizon behaviour leaves $V_+$ positive and monotonically increasing toward $r_{h}$, such states exist and are characterized by complex frequencies $\omega=\omega_{R}+i\omega_{I}$ with $\omega_{I}<0$, the magnitude of which sets the decay time of the corresponding mode. The existence and location of these states are governed by the shape of the plasma distribution: increasing $r_{0}$ broadens the well and shifts the quasibound levels toward lower $\omega_{R}$, whereas increasing $\omega_0$ deepens the well and pushes the levels upward. A systematic determination of the polar quasibound spectrum would require a method that does not rely on the single-barrier approximation---for instance, direct frequency-domain integration of the master equation, a continued-fraction expansion of the radial function, or a Leaver-type continued-fraction scheme adapted to the plasma-corrected potential. We leave such an analysis for future work and report the polar potential here as a benchmark for those numerical approaches.

\begin{table*}[ht]
\centering
\caption{Polar sector: eigenfrequencies for selected values of the parameters in the power-law model with $\ell=0$, obtained with the WKB and Prony methods. Real and imaginary parts are quoted in units of $M^{-1}$.}
\label{table:power2}
\begin{ruledtabular}
\begin{tabular}{|c|c|c c|c c|}
& & $h=1$ & & $h=2$ & \\ [0.6ex] \hline
$r_0$ & $\omega_0$ & WKB & Prony & WKB & Prony \\[0.6ex] \hline
& $0.3$ & $0.203893-0.0679739 i$ & $0.202503-0.063005 i$ & $0.202765-0.0909206 i$ & $0.203277-0.089130 i$ \\
$4$ & $0.5$ & $0.347892-0.06261 i$ & $0.347905-0.062413 i$ & $0.367989-0.0943872 i$ & $0.367943-0.094343 i$ \\
& $0.7$ & $0.491013-0.0625158 i$ & $0.490978-0.062382 i$ & $0.526778-0.0952923 i$ & $0.526780-0.095285 i$ \\ \hline
& $0.3$ & $0.272898-0.0631466 i$ & $0.273374-0.062553 i$ & $0.388039-0.0945601 i$ & $0.388021-0.094570 i$ \\
$7$ & $0.5$ & $0.463509-0.0625219 i$ & $0.463497-0.062394 i$ & $0.663907-0.0956287 i$ & $0.663915-0.095619 i$ \\
& $0.7$ & $0.651802-0.0625028 i$ & $0.651844-0.062347 i$ & $0.936098-0.0959209 i$ & $0.936124-0.095906 i$ \\ \hline
& $0.3$ & $0.329385-0.062654 i$ & $0.329452-0.062479 i$ & $0.566073-0.0954129 i$ & $0.566077-0.095405 i$ \\
$10$ & $0.5$ & $0.555514-0.0625076 i$ & $0.555531-0.062369 i$ & $0.955481-0.095933 i$ & $0.955509-0.095918 i$ \\
& $0.7$ & $0.780125-0.0625008 i$ & $0.780212-0.062347 i$ & $1.34231-0.096076 i$ & $1.342398-0.096051 i$ \\[0.6ex]
\end{tabular}
\end{ruledtabular}
\end{table*}

Among the four plasma profiles, the power-law model is the only one for which the polar potential preserves the single-barrier structure required by both the WKB and Prony methods, and we can therefore extract its eigenfrequencies on the same footing as in the axial sector. The results are collected in Table~\ref{table:power2} for three representative values of the plasma scale length $r_{0}$, three values of the plasma frequency $\omega_0$, and the two indices $h=1$ and $h=2$, with the multipole number fixed at $\ell=0$. Three features stand out. First, the WKB and Prony values agree to better than one percent across the entire parameter space, which confirms that the single-barrier approximation is well satisfied for this profile. Second, increasing $\omega_0$ raises the real part of the frequency almost linearly while leaving the imaginary part essentially unchanged, so the plasma frequency acts primarily as a shift of the oscillation scale rather than as a source of damping. Third, increasing $r_{0}$ has a similar effect on $\omega_{R}$, although the sensitivity is weaker: doubling $r_{0}$ from $4$ to $10$ enhances $\omega_{R}$ by roughly a factor of $1.6$ at $\omega_0=0.5$, whereas the imaginary part remains pinned near $-0.0625$ for $h=1$. The index $h$ has the opposite effect: raising it from $1$ to $2$ leaves $\omega_{R}$ essentially unchanged for $\ell=0$ but increases $|\omega_{I}|$ by about $50\%$, so steeper plasma profiles are more strongly damped. This behaviour reflects the fact that for $\ell=0$ the centrifugal barrier is absent, and the potential is supported entirely by the plasma; the imaginary part of the frequency is therefore directly sensitive to the radial gradient of $\omega_0$, which grows with $h$.

\section{Conclusions}\label{Conclusions}

In this work, we have investigated electromagnetic perturbations in a Schwarzschild spacetime surrounded by a superconducting plasma environment, extending the standard plasma formalism to include the constitutive relation of superconductivity through the covariant London equation. We derived the coupled Maxwell--London system, which takes the form of a Maxwell--Proca equation with the superconducting plasma frequency $\omega_s$ acting as an effective mass for the electromagnetic field. For spatially inhomogeneous plasma distributions, both the plasma frequency and the London penetration depth become position-dependent, leading to a modified wave equation with additional terms arising from the spatial variation of $\omega_s$. We considered five representative plasma density profiles---homogeneous, isothermal gas, exponential, Gaussian, and power-law---each exhibiting distinct asymptotic behaviours.

For the axial (odd-parity) sector, we obtained a single decoupled Schr\"odinger-like wave equation with an effective potential $V_-(r)$ that depends explicitly on the plasma distribution. Our analysis revealed that the asymptotic behaviour of the effective potential is governed by the plasma frequency at large radii: while homogeneous and isothermal distributions yield a constant asymptotic value $\omega_0^2$, the exponential, Gaussian, and power-law models produce potentials that vanish at infinity. We showed that the location of the potential's stationary point depends on the plasma parameters, reducing to the standard Schwarzschild value of $r_0=3M$ only in the vacuum limit. For sufficiently large plasma frequencies, the effective potential may lose its peak structure entirely, leading to the disappearance of discrete quasinormal modes---a cutoff phenomenon that restricts the existence of trapped eigenmodes to specific parameter regimes. Our investigation of the polar effective potential revealed that while it coincides with the axial potential at large radii (as expected from the parity-independence of the local mass term), the two sectors diverge significantly near the horizon. The power-law distribution emerges as the most well-behaved model in the polar sector, maintaining a positive, single-peaked, and well-localized potential across the parameter space considered. In contrast, the Gaussian potential exhibits a steep rise beyond $r\simeq8M$, and the isothermal, exponential, and Gaussian models develop negative dips near the horizon for certain parameter choices.

The polar (even-parity) sector presents a fundamentally different structure. Unlike the axial case, the polar perturbations are governed by a coupled system of equations involving the profile functions $u_{(0)}$, $u_{(1)}$, and $u_{(2)}$. In the vacuum limit, the polar sector separates into physical vector modes and pure gauge scalar modes. However, in the presence of plasma, this separation breaks down, and the coupling between vector and scalar solutions becomes dynamically important. For the monopole mode ($\ell=0$), we derived a single wave equation with an effective potential $V_+(r)$ that includes additional derivative terms proportional to $\omega_s'/\omega_s$ and $\omega_s''/\omega_s$. These terms significantly modify the near-horizon behaviour of the potential, potentially introducing negative regions that signal instabilities or producing steep rises that violate the single-barrier assumption underlying standard WKB analysis.

The numerical analysis of the axial sector, performed using both the sixth-order WKB method with Pad\'e approximation and time-domain integration with Prony fitting, demonstrated excellent agreement between the two techniques, with differences typically of order $10^{-3}\%$. Our results show that the quasinormal frequencies are sensitive to both the plasma frequency $\omega_0$ and the characteristic scale length $r_0$. For the homogeneous plasma, increasing $\omega_0$ increases both the real and imaginary parts of the eigenfrequencies, indicating a destabilizing effect. The isothermal-gas model exhibits similar behaviour with respect to $\omega_0$, but increasing $r_0$ decreases both parts, thereby enhancing stability. The exponential and Gaussian distributions display a different pattern: $r_0$ acts analogously to $\omega_0$, increasing both the real and imaginary parts and thus reducing stability. For the power-law distribution, we found that increasing the fall-off exponent $h$ decreases both the real and imaginary parts, while increasing the combination $\omega_0^2\times r_0^h$ has the opposite effect.

A particularly striking result emerged for the homogeneous plasma configuration: due to its constant asymptotic potential, this model supports quasi-resonances with extremely long lifetimes, where the imaginary part of the frequency is orders of magnitude smaller than in other models. The time-domain evolution reveals an almost non-decaying signal, making the extraction of the damping rate challenging. This behaviour underscores the qualitative distinction between plasma models with constant versus vanishing asymptotic densities.

Due to the pathological features of the polar potential---negative regions, steep rises, and the breakdown of the single-barrier approximation---we did not attempt to extract polar quasinormal frequencies using standard techniques. Instead, we framed the polar eigenvalue problem in terms of quasibound states, which are regular at the horizon and decay exponentially at infinity. The existence and properties of these states are controlled by the plasma distribution: increasing $r_0$ broadens the well and lowers the real part of the frequency, while increasing $\omega_0$ deepens the well and shifts the levels upward. A systematic determination of the polar quasibound spectrum will require numerical methods beyond the WKB approximation, such as direct frequency-domain integration or Leaver-type continued-fraction techniques adapted to the plasma-corrected potential.

Several avenues for future work emerge from this study. First, the extension of our analysis to rotating black holes (Kerr spacetime) would introduce frame-dragging effects that couple the axial and polar sectors, potentially leading to rich phenomenology. Second, the inclusion of self-consistent backreaction of the electromagnetic field on the plasma distribution could reveal feedback mechanisms that modify the spectral properties. Third, the application of our framework to more realistic astrophysical scenarios such as magnetar magnetospheres or accretion disks around black holes would require incorporating additional physical effects such as rotation, magnetic-field geometry, and finite-temperature corrections to the superconducting gap. Finally, the development of robust numerical techniques for the polar sector, capable of handling the complex potential structure we have identified, represents an important direction for future research.

In summary, our work demonstrates that the presence of a superconducting plasma environment around a Schwarzschild black hole significantly modifies the spectrum of electromagnetic perturbations in a manner that depends sensitively on the spatial distribution of the plasma. The axial sector admits a clean spectral analysis with well-defined quasinormal modes, while the polar sector exhibits a richer and more complex structure that challenges standard approximation schemes. These results provide a foundation for understanding electromagnetic wave propagation in exotic plasma environments near compact objects and highlight the importance of considering superconducting phases in astrophysical contexts where extreme densities and strong magnetic fields may favour their formation.

\acknowledgements{This research was funded by the National Natural Science Foundation of China (NSFC) under Grant No.~U2541210.}

\bibliography{References}
\end{document}